\documentclass[twocolumn]{aa}
\usepackage{graphicx}
\usepackage{txfonts}
\usepackage{natbib}
\def\ltsima{$\; \buildrel < \over \sim \;$}
\def\simlt{\lower.5ex\hbox{\ltsima}}
\def\gtsima{$\; \buildrel > \over \sim \;$}
\def\simgt{\lower.5ex\hbox{\gtsima}}
\begin{document}
   \title{The distance of M~33 and the stellar population in its outskirts}

   \author{S. Galleti\inst{1,2},
          M. Bellazzini\inst{1}, F.R. Ferraro\inst{2}
          }

   \offprints{S. Galleti}

   \institute{INAF - Osservatorio Astronomico di Bologna, Via Ranzani 1,
   40127 Bologna, Italy
               \and
             Dipartimento di Astronomia, Universit\`a di Bologna,
              Via Ranzani 1, 40127 Bologna, Italy\\
             \email{silvia.galleti2@unibo.it, michele.bellazzini@bo.astro.it,
	     francesco.ferraro3@unibo.it}}  
   \authorrunning{Galleti, Bellazzini \& Ferraro}
   \titlerunning{The distance and the halo population of M~33}
   \date{Received 22 March 2004 ; accepted 10 May 2004}

   \abstract{We present deep V,I photometry of two $9.4\arcmin\times
   9.4\arcmin$ field in the outer regions of the M~33 galaxy. We obtain a
   robust detection of the luminosity of the Red Giant Branch Tip
   ($I^{TRGB}=20.72\pm 0.08$) from which we derived a new estimate of the
   distance modulus of M~33,  $(m-M)_0=24.64\pm 0.15$, corresponding to a
   distance $D=847\pm 60$ Kpc. By comparison of the color and magnitude of the
   observed Red Giant Branch  stars with ridge lines of template globular
   clusters we obtained the photometric metallicity distribution of the
   considered fields in three different metallicity scales. The derived
   metallicity distributions are very similar over a range of distances from
   the galactic center $10\arcmin\le R\le 33\arcmin$, and are characterized by
   a well defined peak at $[M/H]\simeq-0.7$ ($[Fe/H]\simeq -1.0$, in the Zinn
   \& West scale) and a weak metal-poor tail reaching $[M/H]\simlt -2.0$. 
   Our observations
   demonstrate that Red Giant Branch and Asymptotic Giant Branch 
   stars have a radial
   distribution that is much more extended than the young MS stars associated
   with the star-forming disc. 

}

   \maketitle %

\section{Introduction}

\footnote{Based on observations made with the Italian Telescopio Nazionale
Galileo  (TNG) operated on the island of La Palma by the Centro Galileo Galilei
of  the INAF (Istituto Nazionale di Astrofisica) at the Spanish Observatorio 
del Roque de los Muchachos of the Instituto de Astrofisica de Canarias.} The
accurate determination of the distance to Local Group galaxies is critical  for
the establishment of a reliable extra-galactic distance scale.  In particular,
local late-type galaxies as the  Large Magellanic Cloud, M~31  and M~33 are the
sites of choice for the calibration of several secondary  distance indicators. 

The Sc II-III spiral M~33 (NGC~598 or Triangulum galaxy) is the
third-brightest  member of the Local Group \cite[see][for a review]{syd}.  In
spite of that, while Cepheid variables were  discovered as early as 1920 in
this galaxy \citep{hubble}, reliable distance estimates have become available
only after 1980 \citep{sandage,christian} and differences of the order of 
$\sim 0.3$ mag can be found even among the most recent estimates of the M~33
distance modulus \cite[see][and references therein]{mcc}.

Concerning the stellar populations of M~33, most of the studies have centered
their attention on the massive (young) stars populating the disc  \cite[see,
for example][and references therein]{syd33,syd,urba}, and/or on the innermost
regions of the galaxy \citep{ken,steph}.  On the other hand, only a few authors
have provided some insight into the outer regions and the (possible) halo 
population ( see \citet{mould}, \citet{wilson}, \citet{davidge} and the HST-WFPC2 study by
\citet{kim02}).

Here we present V and I photometry (reaching $V\sim 24.5$) of two fields
located at $\sim 15 \arcmin$ and $\sim 28 \arcmin$ from the center of M~33,  to
the North-West of the main body of the galaxy, approximately in the direction
of the minor axis (see Fig.~1). The Tip of the Red Giant Branch (TRGB) is
cleanly detected with our data and we use it to derive a robust estimate of the
distance modulus of M~33 adopting the RR Lyrae-independent calibration provided
by \citet{tip}, \cite{tip2}. This is the main aim of the present study which is part of
a large programme devoted to the determination of homogeneous distances for 
most of the galaxies of the Local Group \cite[see][]{draco}.  
The  metallicity distribution of the RGB stars in the considered fields
is also obtained from photometric estimates, following the method 
by \citet{m31}. Shortly after this paper was submitted, 
a preprint was posted \cite[][hereafter T04]{t04}
presenting the analysis of a $6.8\arcmin \times 6.8\arcmin$ field located in the
South-Eastern region of M~33 at $\sim 20\arcmin$ from its center. The analysis
is very similar to that performed here and the results are in excellent
agreement, as we will show below.

The structure of the paper is as follows: in Sect.~2 we describe the
observational material, the data reduction process and the photometric
calibration; in Sect.~3 we present the color-magnitude diagrams (CDMs),  our
TRGB estimate of the distance modulus, and the (photometric) metallicity
distributions. The stellar content of the considered fields is also briefly
discussed. A brief summary and the main conclusions are reported in Sect.~4.

\section{Observations and data reduction}

\subsection{Observations}

The observations were carried out on September 7 and 8, 2002, with the
imager/spectrograph DoLoRes at the 3.52m Italian telescope Telescopio Nazionale
Galileo (TNG) in La Palma (Canary Island, Spain). DoLoRes is equipped with a 
2048 x 2048 pixel thinned and back-illuminated Loral CCD array with a total 
field of view of 9'.4 x 9'.4, and a scale of 0.275~"/px.  The seeing was
remarkably good and stable, ranging between  $0.8 \arcsec$ and $1.1 \arcsec$ 
FWHM for all the scientific images.

Two fields were imaged along the minor axes in V and I filters. The locations of
the fields are shown in Figure 1 and listed in Table 1 along with the date, 
the filter and the exposure time of all the scientific exposures.  A total of 7
V and 6 I 600 s exposures have been acquired for the inner field (Field 1,
hereafter F1) and 8 V and 6 I 600 s for the outer field  (Field 2, hereafter
F2). The distance of the field center from the center of the galaxy is $\simeq
15\arcmin$ for F1 and $\simeq 28 \arcmin$ for F2.

   \begin{figure}
   \centering
   \includegraphics[width=9.0cm]{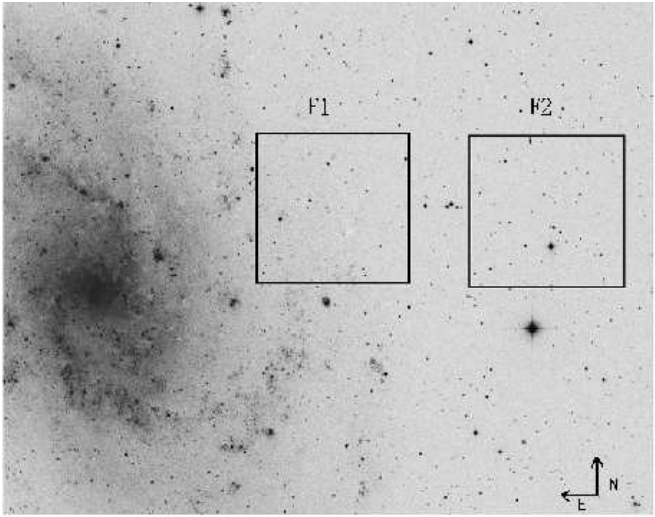}
    \caption{The position of the observed fields over-plotted on a DSS-II 
    image of M33. The dimension of the image is $40\arcmin\times
    33\arcmin$.
    }
              \label{FigMap}%
    \end{figure}


\begin{table}
\caption{Log of the observations}  
{
\begin{tabular}{|lccccc|} 
\hline \hline 
& & & & &\\
Field&RA (J2000) &Dec (J2000)&Date&Filt.&Exp T(s) \\
 \hline 
& &   &	& &	 \\
F1   & 01 32 44.85     & +30 45 02.5& Sept,7,02&V & 3x600 \\
     &                 &            & Sept,7,02& I & 3x600\\
     &                 &            &Sept,8,02 & V & 4x600\\
     &                 &            & Sept,8,02& I & 3x600\\
F2   & 01 31 44.97 & +30 45 00.0    & Sept,7,02& V & 5x600 \\
	&                 &         &Sept,7,02 & I & 4x600\\
     &                 &            &Sept,8,02 & V & 3x600\\
     &                 &            & Sept,8,02& I & 2x600\\
	   \hline \hline  
\end{tabular}
}
\end{table}

\subsection {Data analysis} All the images were corrected for bias and
flat-field  using  standard IRAF procedures. The relative photometry was
carried out with the PSF-fitting code DoPhot (\cite{dophot}). Each frame was
independently reduced. A 3 $\sigma$ threshold above the background noise was
adopted for the search for sources on the frames and the spatial variations of
the PSF were  modeled with a quadratic polynomial function.  Only the sources
classified as stars by the code were retained. All the V and I catalogues of a
given field were reported to the  (instrumental) photometric system of the
best-seeing image acquired under photometric conditions. 
The catalogues were then cross-correlated, the magnitudes were
averaged and the standard deviation adopted as the photometric
uncertainty of the individual stars.  In the final catalogue of instrumental
magnitudes and positions we have retained only the sources that have at least
three valid measures of the magnitude for each passband (e.g. at least 3 V and
3 I). Moreover, all the stars with an associated error (either in V or I
magnitude) larger than three times the average uncertainty at their magnitude
level were excluded from the sample. The final F1 and F2 catalogues contains 
26399 and 1937 sources, respectively. Accurate aperture corrections have been 
obtained for each field on a few tens of bright and isolated stars.

The stellar crowding is quite low in the considered fields that sample 
external low-surface brightness regions of the galaxy. Even in F1 
the average stellar density is as  low as $0.08~stars/arcsec^2$ 
(considering {\em all} the detected stars). 
For the
applications presented in this paper the effects of incompleteness are not an
issue since we always compare subsets of stars that are homogeneous in magnitude
and color (see Sect.~3.3 and 3.4 below). No attempt is made to determine the
{\em true} Luminosity Functions of the observed sequences.
The TRGB level, the main target of the
present study, occurs more than 3 magnitudes above the limiting magnitude, a
range in which the considered sample is likely $\sim 100$\% complete. Since the
photometric uncertainties of individual stars (in each passband) are 
empirically estimated as the standard deviations of 3 to 8 independent measures
of the magnitudes there is no need for artificial star experiments to
characterize photometric errors.

\subsection {Photometric calibration}

The absolute calibration has been obtained from several repeated observations 
of \cite{landolt} standard fields, including all the stars listed in the 
extended catalog of calibrators provided by \cite{stetson}.  The coefficients
of atmospheric extinction ($C_{ext}$) were directly obtained  by repeated
observations of the same standard field at different air mass.  The final
calibrating equations and $C_{ext}$ are shown in the upper panels of Figure 2.

To check the accuracy (and reproducibility) of the photometric calibration we
reduced one V and one I 30 s frames centered on the globular cluster NGC 6779,
acquired during the second night of the run. The catalogue was calibrated with
the relations shown in Fig.~2 and the final photometry was compared with the
 photometry of the cluster provided by \cite{ros}. The results of
the comparison (shown in the lower panels of Fig.~2) demonstrate that our
photometry is in excellent agreement with that of \cite{ros}, an the
characteristic uncertainty of our absolute calibration is of the order of 
$\le \pm 0.02$ mag. 


   \begin{figure}
   \centering
   \includegraphics[width=9.3cm]{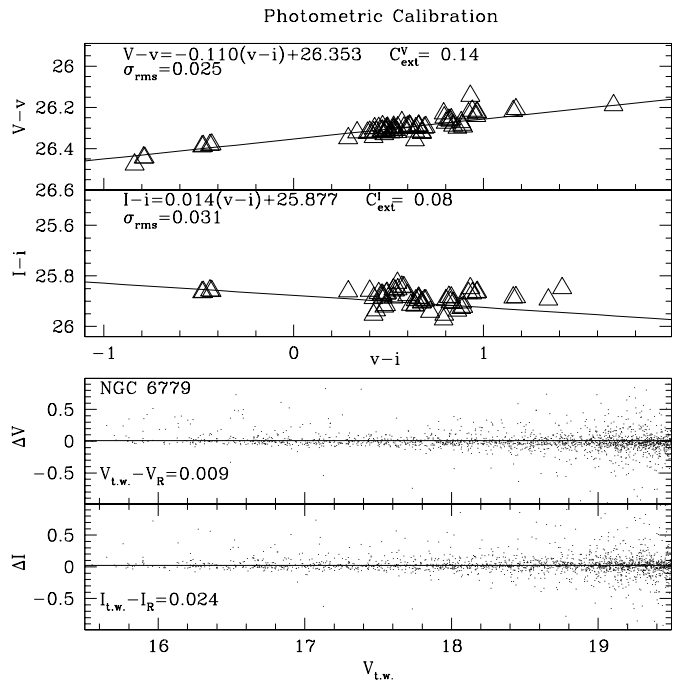}
    \caption{Upper panels: difference between tabulated magnitudes (V,I)
    and instrumental magnitudes (v,i) vs instrumental color index 
    (v-i) for the observed \cite{landolt} standard stars. 
    The calibrating relations are plotted (solid lines) and the
    corresponding equations are reported along with the RMS of the linear fits
    and the value of extintions coefficients.
    Lower panels: comparison of the photometry of NGC~6779, obtained from data
    acquired during the same run of the M~33 data and calibrated with the
    relations above, and the independent photometry by \cite{ros}. 
}
              \label{FigCal}%
    \end{figure}


\section{Results}

\subsection{Color-Magnitude Diagram}

The Color Magnitude Diagrams (CMD) of the observed field are presented in
Fig.~3.  The CMD of F1 is dominated by a wide RGB sequence, running from $I\sim
20.6$ down to the limiting magnitude of the photometry. The exact location of
the TRGB (as derived in Sect.~3.2, below) is reported in the CMDs as a
horizontal line to the red of the observed RGB.   The sources brighter and
redder than the RGB Tip are likely bright Asymptotic Giant Branch (AGB) stars,
probably associated with an intermediate-age (and/or metal-rich) population
\cite[see][for an extensive discussion]{davidge}. The presence of a
conspicuous plume of intermediate-young Main Sequence (MS) stars reaching 
$I\simeq 19.4$ is also evident on the blue side of the CMD, 
with a sharp blue edge at $V-I\simeq 0.0$.  In
the CMD of F2 only the RGB and AGB population are discernible, while the blue
MS stars are completely absent. Star
counts on the RGB indicate that the stellar density drops by a factor $> 20$
from F1 to F2.

The contamination by foreground Galactic stars is negligible for the purposes
of the present study. The Galactic model by \citet{robin} predicts fewer than
200 Galactic stars in the observed fields within the color and magnitude ranges
spanned by the CMDs of Fig.~3. By inspection of the synthetic CMD obtained with
the \citet{robin} model we conclude that even in the poorly populated F2 the
only region of the CMD that may be significantly affected by Galactic
contamination is that which hosts bright AGB stars, e.g. $V-I\simgt1.8$ and 
$V\simlt 20.6$ (see Sect.~3.4, below).

   \begin{figure}
   \centering
   \includegraphics[width=9.3cm]{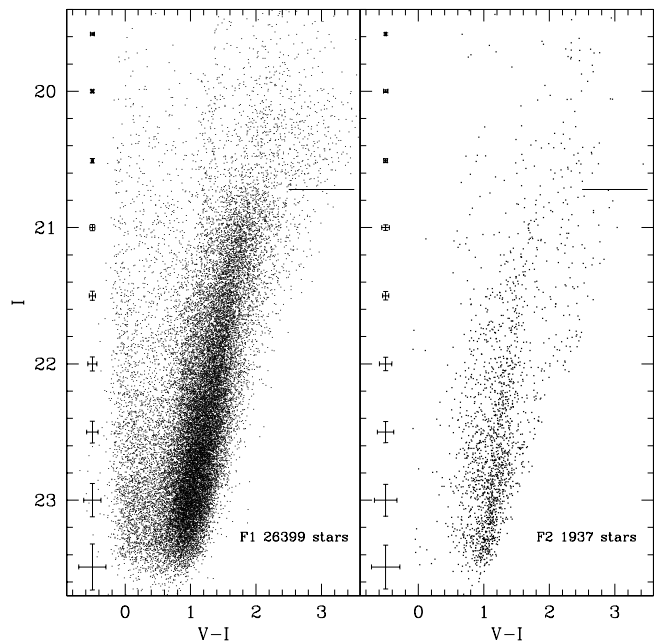}
     \caption{Color Magnitude Diagrams of the two observed fields. The errobars
     show the average photometric uncertainty as a function of magnitude; the
     horizontal lines marks the position of the TRGB as determined in Sect.~3.2.
               }
          \label{FigCM}
   \end{figure}


\subsection {TRGB distance}

The use of Tip of the Red Giant Branch (TRGB) as a standard candle is now a
widely used technique to estimate the distance to galaxies of any
morphological type  \cite[see][for a detailed description of the method, recent
reviews and applications]{lfm93,mf95,mf98,walk}. The underlying physics is well
understood \citep{mf98,scw} and the observational procedure is operationally
well defined \citep{mf95}. The key observable is the sharp cut-off occurring at
the bright end of the RGB Luminosity Function (LF) that can be easily detected
with the application of an edge-detector filter \citep[Sobel
filter,][]{mf95,smf96} or by other (generally parametric) techniques \cite[see,
for example][]{mendez,mcc}. The necessary condition for a safe application of
the technique is that the observed RGB Luminosity Function should be well
populated, with more than $\sim 100$  stars within 1 mag from the TRGB
\citep{mf95,draco}. 

The F2 sample is not sufficiently populated for a safe application of  the
method while the F1 sample clearly fulfils the above criterion (there are 
more than 2500
RGB  stars within 1 mag from the TRGB), hence we limit the TRGB research to F1.
As a first step, to limit the range of metallicity of
the stars involved in the TRGB detection, we select RGB stars  by color
following the approach adopted by \citet{mcc}. The adopted selection includes
the main bulk of the RGB population and it is shown
in the lower left panel of Fig.~4. The logarithmic LF is presented as an ordinary
histogram and as a {\em generalized histogram} \cite[e.g. the  histogram
convolved with a Gaussian with standard deviation equal to the photometric
error at the given magnitude, see][for definitions  and
references]{laird,draco} in the upper left and upper right panels of Fig.~4,
respectively. The sharp cut-off is an obvious feature of both representations
of the LF and is easily detected by the Sobel filter (Fig.~4, lower right
panel). As usual, the peak of the filter response is taken as the best estimate 
of the
TRGB location and the Half Width at Half Maximum of the same peak is taken as
the associated uncertainty, $I^{TRGB}=20.72\pm 0.08$. If we consider the most
recent estimates in the literature, our value is $\sim 2.2 \sigma$ larger than
that found by \citet[][$I^{TRGB}=20.54\pm0.01$; but these authors  provide only
a formal error on their estimate]{mcc}, and $\sim 1 - 2 \sigma$  lower than the
estimates by \citet[][$I^{TRGB}=20.82 - 20.92 \pm0.05$ depending  on the
considered field]{kim02}, e.g. it is bracketed by the two quoted results.
On the other hand our estimate is in excellent agreement with that 
obtained by T04 ($I^{TRGB}=20.75\pm0.02$). 

We adopt $E(B-V)=0.04$, according to the reddening maps by \citet{cobe} and
\citet{bur} and $A_I=1.76E(B-V)$, according to \citet{dean}.  We note however
that most of the other available estimates of the foreground reddening cluster
around $E(B-V)=0.08$ \cite[see][and references therein]{syd33}. To account for
this, we report also the results we obtain adopting $E(B-V)=0.08$ (see Table~2,
below). Note that the effects of this different assumption are small either on
the final distance modulus (e.g. $0.05$ mag) or the average metallicity
($\le 0.15$ dex; see Tab.~2). 
According to the detailed dust maps of M~33 by \citet{hippe}
the effect of the intrinsic extinction  should be negligible in the fields
considered here. 

In \cite{tip} we have provided a robust zero-point to the calibrating relation
providing the absolute I magnitude of the tip ($M^{TRGB}_I$) as a function of
metallicity  ($[Fe/H]$, in the \citet{zw84} scale, hereafter ZW), based on the
geometric distance to the cluster $\omega$ Centauri obtained by \citet{ogle}
using the double-lined detached eclipsing binary OGLE-17. This calibration is
fully independent of the usual (Cepheid based and/or RR Lyrae based) distance
scales. In \citet{tip2} we have extended the calibration to Near Infrared
passbands and we refined our I calibration providing also the relation for
$M^{TRGB}_I$ as a function of the {\em global metallicity} ([M/H], see
\citet{scs93} and \citet{f99} for definitions and discussion) that we adopt in
the present analysis.

   \begin{figure}
   \centering
   \includegraphics[width=9.3cm]{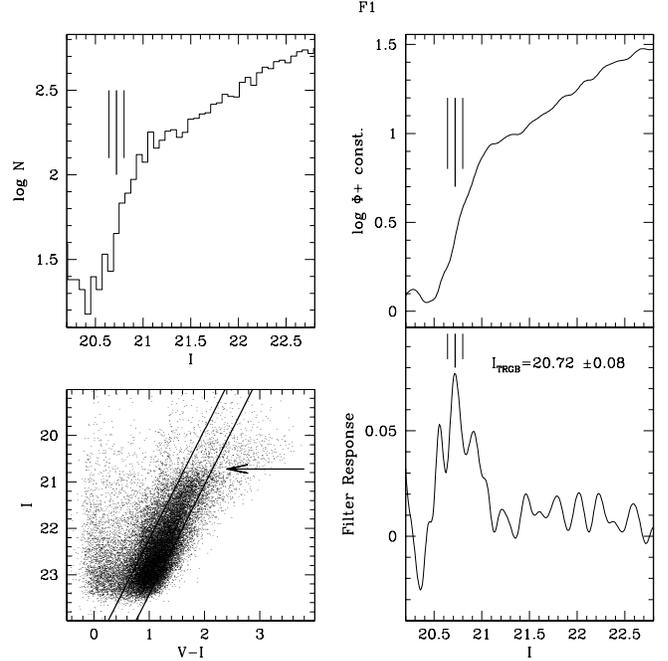}
      \caption{Detection of the TRGB. The CMD in the lower left panel shows the
      adopted selection, e.g. the stars enclosed by the two diagonal lines.
      The arrow marks the position of the TRGB.
      The upper panels display the logarithmic LF of the upper RGB as an
      ordinary histogram (left) and as a generalized histogram (right). 
      The thick lines marks the position of the TRGB, the thin lines enclose 
      the associate uncertainty range. Lower panel: Response of the Sobel's
      filter to the observed LF.}
          \label{FigTip}
  \end{figure}

Since the distance modulus derived from $I^{TRGB}$ is weakly dependent on
metallicity, and our metallicity estimates (obtained by comparison with
template RGB ridge lines, see below) depend on the assumed distance modulus, we
adopted  an iterative method to find simultaneously the two 
quantities searched for.  First we derived a preliminary distance modulus adopting 
$M^{TRGB}_I=-4.04$, then we derived a median metallicity of the considered
population as described in Sect.~3.3 below, and we obtained a refined estimate 
of the modulus using the obtained median metallicity ($[M/H]=-0.75$, see below)
as an input for the calibrating relation by \cite{tip2}:

\begin{equation} M^{TRGB}_I=0.258[M/H]^2+0.676[M/H]-3.629  ~~~(\pm 0.12).
\end{equation}

The process converged to stable values of the distance modulus and of the
median metallicity in 2-3 iterations, independently of the assumed reddening
and/or the considered metallicity scale. Our final estimate (for $E(B-V)=0.04$
and $[M/H]_{med}=-0.75$) is $(m-M)_0=24.64\pm 0.15$, where all the sources of
uncertainty have been taken into account. The corresponding distance estimate
is $D=847\pm 60$ Kpc.

\subsection {Metallicity}

We derive the Metallicity Distributions (MD) of the studied fields from the
Color and Magnitude distribution of RGB stars (transformed to the absolute
$(V-I)_0$ vs. $M_I$ plane adopting the reddening and distance modulus described
above) by interpolation on a grid of RGB ridgelines of template globular
clusters, adopting the same scheme as \citet{m31}. 
Essentially the same approach is adopted also by T04.
The MDs are obtained in
different metallicity scales, e.g. the ZW scale, the scale by 
\citet[][hereafter CG]{cg} and the global metallicity scale described in
\citet{f99}, to make the comparison with other studies easier. While the
individual photometric metallicities  provided by the adopted procedure may be
quite uncertain, the overall metallicity distribution and its average
properties are sufficiently well characterized to provide interesting insights
into the stellar population under analysis and it has been widely used, in recent
years, in the study of resolved  galaxies \cite[see][for details, discussion
and references]{m31}. The color distribution of RGB stars should 
depend - to a lesser extent - also on the {\em age} distribution of the
underlying population. T04 studied this problem in the case of M~33 by mean of
synthetic CMDs drawn from theoretical evolutionary tracks and concluded that
``...the ages of the RGB stars are not likely to significantly affect the
derived MDs...''. They estimate that the maximum expected shift of the peak of
the MDs is of the order of 0.1-0.2 dex. Finally, the young MS population observed in 
F1 - and not in F2 - cannot affect the comparison between the MDs of the two 
fields since their evolved counterparts should be negligible in number and
do not fall in the selection box we adopt to obtain MDs (see Fig.~5 and 7,
below). When dealing with photometric metallicities it should 
be kept in mind that the underlying age distribution may affect the derived
MDs.

In Fig.~5 a direct comparison between the observed RGBs of F1 and F2 and the
adopted template ridge lines is presented.  It is immediately clear that the
large majority of M~33 RGB stars (in both fields) are enclosed within the ridge
lines of M~5 ($[Fe/H]_{CG}=-1.11$; $[M/H]= -0.90$) and of 47~Tuc 
($[Fe/H]_{CG}=-0.70$; $[M/H]= -0.60$). In the CMD of F2 an 
anomalous clustering of stars can be noted around the ridge line of NGC~6553 
($[Fe/H]_{CG}=-0.16$; $[M/H]= -0.06$) that has no counterpart in the CMD of F1.
However, as can be appreciated from Fig.~3, such red stars are just above the
limiting magnitude at their color. For this reason we will not discuss in detail
this feature in the following. Deeper photometry is needed to firmly assess
the possible presence of an excess of very red RGB stars in this region.

   \begin{figure}
  \centering
   \includegraphics[width=9.3cm]{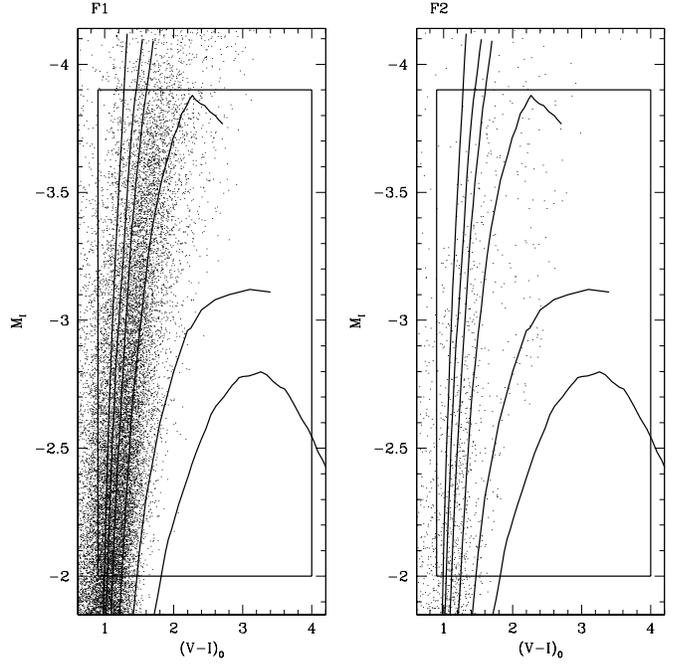}
     \caption{The CMDs of the RGB stars in F1 (left panel) and F2 (right panel)
      are compared to the grid of RGB ridge lines of template globular clusters
      we adopted to derive the metallicity distributions. From blue to red, the
      template clusters are: M~92, M~13, M~5, 47~Tuc, NGC~6553 and NGC~6528.
      The thick box encloses the RGB stars actually selected for the derivation
      of the metallicity distribution \cite[see][for details]{m31}. 
               }
          \label{FigInt}
  \end{figure}
%

   \begin{figure}
  \centering
   \includegraphics[width=9.3cm]{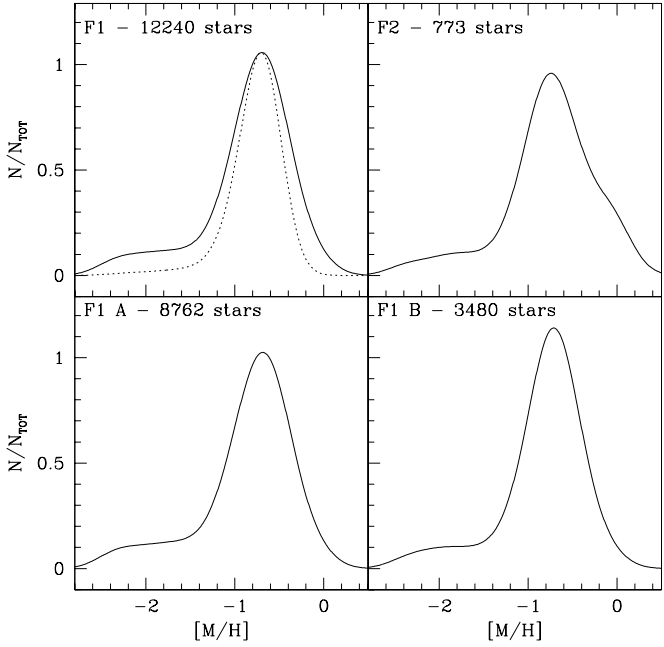}
     \caption{Metallicity Distributions (continuous lines) of the F1 field as a
     whole (upper left panel), of F2 (upper right panel), of the inner part of
     F1 ($10\arcmin \simlt R \simlt 15\arcmin$; F1 A, lower left panel) and of the 
     outer part of the same field ($15\arcmin \simlt R \simlt 20\arcmin$;
     F1 B, lower right panel). The dotted line in the upper left panel shows the
     instrumental response of the method, the MD that would be obtained if the
     width of the RGB was entirely due to the photometric errors.
               }
          \label{FigDMet}
  \end{figure}
%

In Fig.~6 we present the MDs (in the form of generalized histograms) as a function 
of the global metallicity for F1 (upper left panel) and F2 (upper right panel).
To study in finer detail the radial behavior of he MD we split F1 in
two subregions (A and B) of similar area. In particular F1 A contains all the F1
stars less distant than $15\arcmin$ from the center of M~33, while the F1 stars
with $r\ge15\arcmin$  are assigned to F1 B. The MDs of F1 A and F1 B are plotted
in the lower left and lower right panels of Fig.~6, respectively.
Note that if we exclude from the interpolation the stars with $M_I\ge -2.5$,
i.e. in the range where the sensitivity of color to metallicity is lower,  
the obtained MDs are unchanged. This experiment also demonstrates that the
derived MD are not sensitive to the effect of incompleteness, as expected.
The average properties of all the considered MDs are summarized in Table 2, for
two different assumptions of E(B-V). 
There are a number of
considerations emerging from the inspection of Fig.~6 and Table~2:

\begin{enumerate}

\item{} The MD of all the considered fields shows a strong peak at $[M/H]\simeq
-0.7$. This justifies our assumption of the median metallicity as the
characteristic value of the dominant population in our determination of the TRGB
distance. A sparsely populated tail of metal-poor stars extending 
to $[M/H]<-2.0$ is also
present in all the presented MDs. This general similarity over large areas of
the galaxy (F1 covers a range of galactocentric distances from $\sim 2.4$ Kpc
to $\sim 5$ Kpc, F2 from $\sim 5.5$ Kpc to $\sim 8.2$ Kpc) is in agreement with
the results by \citet{kim02} and is reminiscent of what is observed in M~31
\citep{m31}. Our MDs are very similar to those obtained by T04.

\item{} The bell-shaped curve plotted as a dotted line in the upper left panel
of  Fig.~6 displays the response of the adopted interpolation scheme to a 
Simple
Stellar Population \cite[SSP, i.e. a population of stars having the same age
and chemical composition,][]{rf88} observed under the same conditions as our real
data. It has been obtained adopting, as input for the interpolation, 
a synthetic RGB population whose color width is entirely due to the photometric
errors. The I magnitude of the ``synthetic'' stars is extracted from the 
observed RGB luminosity function, the V-I color is obtained from the average
ridge line of the observed RGB plus a photometric error drawn at random from a
Gaussian distribution having $\sigma$ equal to the average {\em observed}
photometric uncertainty at the considered magnitude \cite[as done in][]{draco}. 

To obtain the {\em true}
width of the underlying MD the described ``instrumental response'' should be
deconvolved from the observed MD (continuous line). The main peak of the
observed MD is well fitted by a Gaussian distribution with $\sigma=0.34$ dex,
while the instrumental response curve is well approximated by a Gaussian
distribution with  $\sigma=0.23$ dex. It may be concluded that the {\em true}
intrinsic  dispersion of the main peak of the MD is $\sigma\simeq 0.25$ dex.

\item{} By comparison with ridge lines of template globular clusters,
\citet{mould} estimated $<[Fe/H]_{ZW}>\simeq -2.2\pm 0.8$ for a field at a
galactocentric distance similar to F2. On the other hand, we find, for both F1
and F2, $<~[Fe/H]_{ZW}>\simeq -1.03\pm 0.40$, in excellent agreement with the
results by \citet{davidge}, \citet{cuil} and T04.  
The difference is partly  justified by the
different assumptions about distance  ($(m-M_0)=24.8$ instead of our
$(m-M_0)=24.64$). However, even adopting their distance modulus we find
$<[Fe/H]_{ZW}>\simeq -1.22\pm 0.40$, much more metal-rich than what was found by
\citet{mould}. We tentatively ascribe this difference to a possible problem in
the absolute calibration of Mould \& Kristian's  photometry.  This
hypothesis is confirmed by the results of T04.

\item{} Assuming the same distance modulus as \citet{mould}, \citet{kim02}
obtain $-0.61\le <[Fe/H]_{ZW}>\le -0.86$ for 10 fields covering
(approximately) the same radial range than our F1. The agreement with our
results is much better than with Mould \& Kristian's one, 
still the difference is not
negligible and it is not justified by the different distance modulus assumed (a
larger distance modulus should imply a brighter and hence more metal-poor RGB).
However \citet{kim02} derived their mean metallicity from the mean $(V-I)_0$ 
color at $M_I=-3.5$. We consider our median/mean metallicities, 
derived from {\em all} the RGB stars brighter than $M_I=-2.0$ and based on an 
accurately checked photometric calibration, as more robust and safer than
 those by \citet{kim02}.

\item{} \citet{kim02} found a weak radial gradient in the mean metallicity (0.2
dex) in the range $1\le R\le 5$ Kpc (see also T04). 
We find no sign of variation of the mean
metallicity in the range $2.4 \le R\le 8.2$ Kpc, but such feeble differences
may have gone undetected at the level of accuracy of our relative photometry
(e.g. may be hidden in the ``instrumental width'' of the observed sequences).

\item{} The weak shoulder at $[M/H]\sim -0.2$ in the MD of F2 is due to the
handful of red stars around the ridge line of NGC~6553 discussed above, hence,
at the present stage, cannot be trusted as a real feature of the MD (but it
deserves further investigation). 
\end{enumerate}

   \begin{figure}
  \centering
   \includegraphics[width=9.3cm]{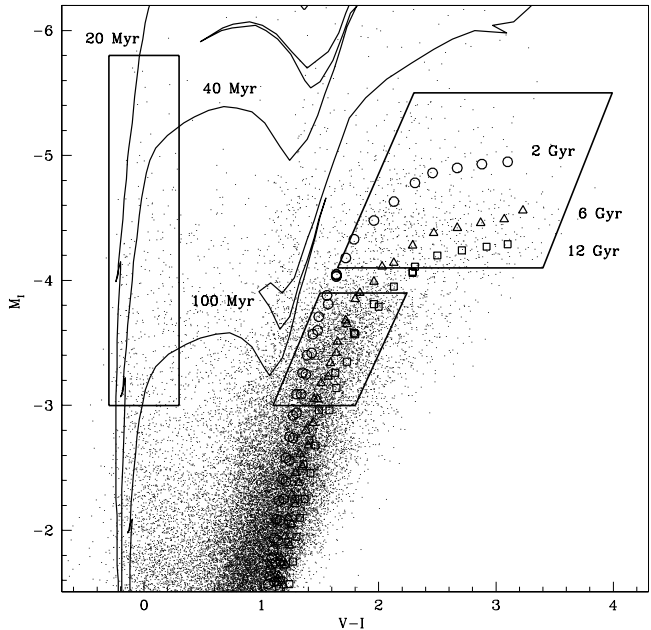}
     \caption{Description of the adopted selection boxes. The box around
     $V-I=0.0$ selects a sample of young MS stars, the large box above the TRGB
     selects AGB stars, the small box below the TRGB selects RGB stars.
     The continuous lines are isochrones of solar metallicity, the open symbols
     are isochrones at $[M/H]=-0.7$, from the set by \citet{girardi}. The ages
     are reported in the plot. The small dots are the observed stars of F1.
               }
          \label{FigBox}
  \end{figure}
%

\begin{table}
\begin{center}
\caption{Distance Modulus, median Metallicity and standard deviations in
different metallicity scales, and
with different assumptions on the reddening. Note that the median and the
mean metallicity are nearly coincident in all the considered cases.}  
{

\begin{tabular}{|l|rr|rr|} 
\hline \hline 

& & & &\\
 &   F1  &  F2   &   F1 &   F2     \\
 \hline 
E(B-V)         &\multicolumn{2}{|c|}{ 0.04 } & \multicolumn{2}{|c|}{ 0.08 } \\
\hline
& &   &	&	 \\
$[Fe/H]_{ZW}$   & -1.03 &-1.03 & -1.17 & -1.18 \\
$\sigma_{ZW}$   &  0.40 & 0.40 &  0.40 &  0.44  \\
$[Fe/H]_{CG}$   & -0.89 &-0.89 & -0.97 & -0.98  \\
$\sigma_{CG}$   &  0.27 & 0.28 &  0.28 &  0.32   \\
$[M/H]$         & -0.75 &-0.74 & -0.81 & -0.81   \\
$\sigma_{[M/H]}$&  0.23 & 0.23 &  0.23 &  0.27   \\
\hline & &   &	&	  \\
$(m-M)_0$      &\multicolumn{2}{|c|}{24.64$\pm0.15$ }&\multicolumn{2}{|c|}{24.59$\pm0.15$ }\\
	   \hline \hline  
\end{tabular}
}
{ 
\begin{flushleft}
\end{flushleft}
}
\end{center}
\end{table}

\subsection{Stellar Populations and radial distributions}

The stellar content of the outer region of M~33 is poorly explored. It is
interesting to investigate if the different kind of stars identified in  our
CMDs share the same spatial distribution. To check this point we defined three
selection boxes that are depicted in the F1 CMD shown in Fig.~7.  Note
that the boxes are defined in a range of magnitude in which incompleteness
effects should be weak or negligible ($M_I<-3.0$, e.g. more than 2 mag
above the limiting magnitude) and cover similar magnitude ranges.

As a
guideline, we have superimposed three isochrones of solar metallicity and age
$=20, 40, 100$ Myr (continuous lines) and three isochrones with $[M/H]=-0.7$ 
and age $=2, 6, 12$ Gyr (empty symbols) from the set by \citet{girardi}. The
bluest box samples the upper MS, e.g. young stars with age $\le 100$ Myr. The
large box above the TRGB samples the bright AGB stars, the smaller box samples
the brightest RGB stars. Both kinds of tracers are associated with intermediate
to old age stars, but are not necessarily linked thogether.  In
particular, bright RGB stars trace populations older than $1-2$ gyr
\cite[see][and references therein]{scw}. 
In the
following, R must be intended as the projected angular distance from the
center of the galaxy.

In the upper panels of Fig.~8 the adopted selection boxes are superposed on the
CMDs of F1 (left panel) and F2 (right panel). The lower panels of Fig.~8 show
the cumulative radial distributions of the stars falling in the boxes in the
two different fields. In the radial range covered by F1 the distributions of
RGB and AGB stars are indistinguishable. On the other hand, MS stars appear
much more centrally concentrated since their distribution seems to end at  $R\sim
17\arcmin-18\arcmin$, e.g. around 2 disc scale-lengths \citep{syd33}. According
to a Kolmogorov-Smirnov test, the probability that the MS and RGB  samples are
drawn from the same parent population is $P\le 0.04$\%. This suggests that the
RGB and AGB stars are not associated with the young disc component traced by MS
stars. At least a (significant) fraction of them should belong to a more
extended galactic component.

The radial distribution of MS stars is not reported in the lower right panel of
Fig.~8 since, in agreement with the above conclusion, F2 is virtually  devoid
of stars populating the MS box. It is surprising to note that in this field, AGB
stars appear to follow a significantly different distribution with  respect to
RGB stars. The distribution of AGB stars is less centrally  concentrated and it
is quite similar to uniform distribution on the sky. The latter fact would
be naturally explained if the AGB sample of F2 would be dominated by foreground
contamination \cite[see also][]{davidge}.  According to the predictions of the
Galactic model by \citet{robin} this seems to be the actual case. The model
predicts that the number of Galactic stars falling in the AGB box is 48, less
than compared to
the 61 actually observed. Hence $\simeq 80$ \% of the putative AGB stars in F2
are likely foreground stars. The impact is much smaller on the F1 stars where
584 stars are observed in the AGB box, hence the fraction of foreground
contaminants is $< 10$ \%. On the other hand the expected number of foreground
stars falling in the RGB box is $\sim 4$, e.g. negligible in both fields.

Therefore, the difference of radial distribution shown in the lower right
panel of Fig.~8 is completely spurious. On the other hand, if we consider star
counts and take into account the corrections for foreground contamination, it
turns out that while the number of RGB stars drops by a factor $\simeq 27$
going from F1 to F2, the number of AGB stars decreases by a larger factor,
$\simeq 42$. This suggests that RGB stars may follow a more extended
distribution with respect to AGB stars at large radii. 

All the above considerations seem to indicate that populations of different
characteristic ages follow different distributions on the sampled scales,  the
older stars having more extended distributions. This is suggestive of the
 presence of a weak ``classical'' old halo component in M~33. The only
previous indication in this sense (from field stars) is provided by the
discovery of a few  candidate RR Lyrae variables by \citet{prit}, while all
other hints of the existence of an old spheroidal stellar component come from
the study of globular clusters \citep{schom,sarajedini}.
 Note, however, that an extended and old disc component is also compatible
with our observations (see T04).

In the upper panel of Fig.~9 we report the radial profiles of RGB and MS stars
over the range(s) sampled by the present study. The MS profile is reasonably
reproduced by an exponential law with scale-length $h=9.2\arcmin$ up to 
$R\simeq 17\arcmin$, in agreement with the results by \citet{kent}. However, at
$R=19\arcmin$ the sharp drop of the density already observed and discussed in
Fig.~8 is clearly evident. On the other hand the RGB profile is well fitted by
an exponential law with $h=4.9\arcmin$ over the whole radial range sampled by
F1. The density of RGB stars falls significantly below the adopted exponential
profile in F2, suggesting a break in the observed profile in the range
$20\arcmin \la R \la 25\arcmin$. In our view, the most interesting result of
this comparison is that, even ignoring the observed density 
cut-offs, the RGB and MS distributions do have significantly different profiles,
again suggesting a different origin.

The lower panel of Fig.~9 shows that in the radial range covered by F1
($10\arcmin\le R \le 20\arcmin$) the observed density profile of RGB stars is
equally well fitted by the exponential law described above and by a
$R^{1\over{4}}$ law having effective radius $R_e=2.7\arcmin$ (in fact, the
$R^{1\over{4}}$ law provides a marginally better fit with respect to the
exponential). It is interesting to note that the same $R^{1\over{4}}$ law was
found to provide a good fit also to the central bulge of M~33
\citep{boule,bothun}, suggesting a possible connection between the bulge and
the putative halo component \cite[but see][for a detailed decomposition of the
inner profile]{steph}. While suggestive, the above result is limited to the considered radial range
($10\arcmin\le R \le 20\arcmin$) where the contribution of disc stars to the RGB
population may be low. To correctly disentangle the contribution of the
disc from that of the more extended component identified here, a complete
sampling of the density profile from the center to the outskirts of the galaxy
is needed, e.g. covering also the regions in which the surface brightness should
be dominated by the exponential disc. This kind of analysis is clearly beyond
the reach of the present study.

   \begin{figure}
  \centering
   \includegraphics[width=9.3cm]{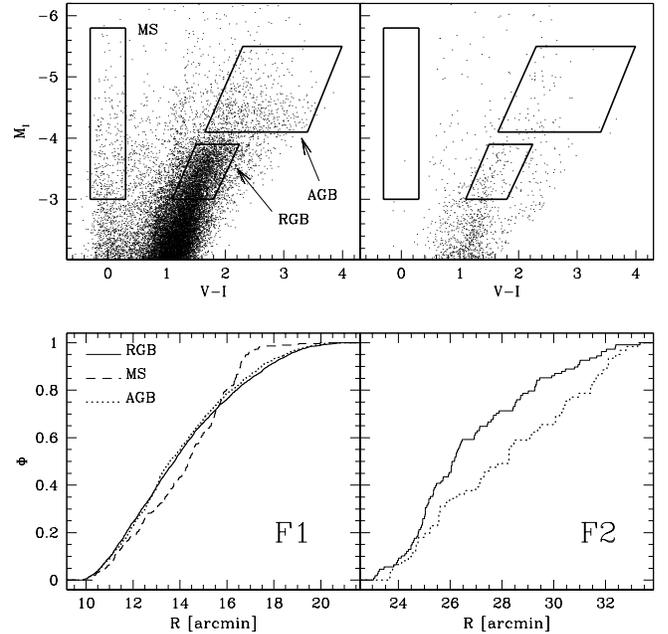}
     \caption{Upper panels: the selection boxes described in Fig.~7 are
     superposed on the CMDs of F1 (left) and F2 (right). The corresponding
     cumulative radial distributions are displayed in the lower panels.
               }
          \label{FigRad}
  \end{figure}
%

   \begin{figure}
  \centering
   \includegraphics[width=9.3cm]{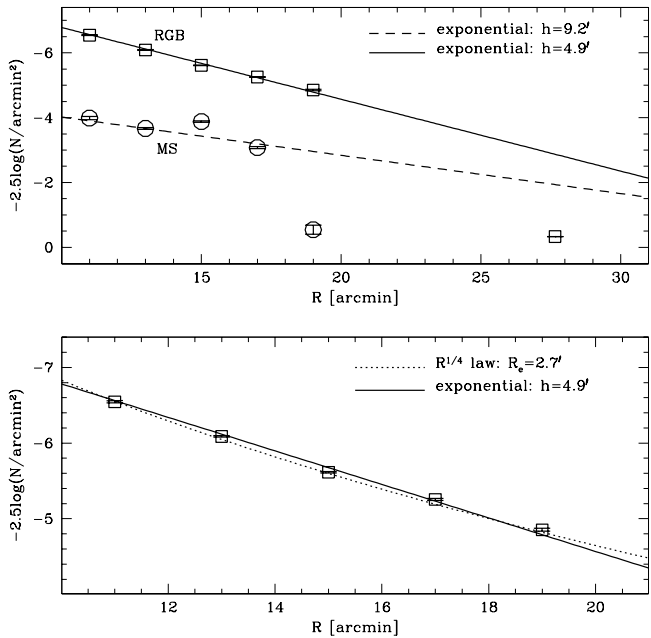}
     \caption{Upper panel: stellar density profiles for MS (open circles) and
     RGB stars (open squares) compared with two different exponential laws.
     All the estimates with $R\le 20\arcmin$ are computed over sections of
     concentric annuli $2 \arcmin$ wide. The RGB points at $R=27.6 \arcmin$ is
     the density estimate obtained from F2 as a whole.
     Lower panel: expansion of the RGB density profile. The continuous line
     is the same exponential law displayed in the upper panel, the dotted line
     is a $R^{1\over{4}}$ law with $R_e=9.2\arcmin$. The associated uncertainty
     is the Poisson noise of the star counts propagated to the adopted unities.
               }
          \label{FigRad2}
  \end{figure}
%

\section{Conclusions}

We have obtained a robust detection of the I magnitude of the TRGB in a field
located at $\sim 15\arcmin$ from the center of M~33, near the galaxy minor
axis.  Adopting the median metallicity we derived from the same data and the
calibration of $M^{TRGB}_I$ as a function of the global metallicity ($[M/H]$)
provided by \citet{tip2}, we have obtained a new estimate of the distance
modulus of M~33, $(m-M)_0=24.64\pm 0.15$. All the sources of uncertainty have
been taken  into account in the reported error bar.

In Fig.~10, our distance modulus is compared with previous estimates available 
in the literature. The large majority of  the reported values are  compatible,
within the formal 1-$\sigma$ errors, with our estimate. The only exceptions are
provided by the oldest analysis of Cepheid variables, based on photographic
plates \citep{madore,christian}. If we exclude these two estimates as well as
those by \citet{argon} and \citet{gree93}, that are affected by very large
uncertainties, we obtain, from 19 independent estimates including ours, an
average distance modulus $(m-M)_0=24.69 \pm 0.15$ (average $\pm$
standard deviation) in excellent agreement with our result. It is interesting
to note that the standard error on the above average is just
$\epsilon=\sigma/\sqrt{19}=0.03$ mag. 

The photometric metallicity distributions described in Sect.~3.3 indicate
that (a) the observed RGB population in the outskirts of M~33 has a typical
metallicity ($[Fe/H]_{ZW}\simeq -1.0$) that is intermediate between  that of
the halo of the  Milky Way ($[Fe/H]_{ZW}\simeq -1.5$) and that of M~31
($[Fe/H]_{ZW}\simeq -0.6$) \cite[see][for discussion and references]{m31}; (b)
the MDs are quite similar everywhere, within the sampled regions, similar to
the case of M~31 \citep{m31}. 

In the radial range $10\arcmin \le R \le 20\arcmin$, the AGB and RGB stars have
a similar radial distribution, much more extended than that of young MS stars
which appear to decrease abruptly around $R=18\arcmin$. The distribution of RGB
stars is equally well fitted by an exponential law (but not the same that fits
the distribution of MS stars) or by the same $R^{1/4}$ law that fits the
central bulge \citep{boule}. The density of RGB stars is observed to fall far
below that predicted by both the adopted best-fit models at $R\sim 27\arcmin$,
but the actual break may occur anywhere between $R=20\arcmin$ and $R=27
\arcmin$ (e.g. between 4 and $\sim 5$ exponential scalelenghts, in good
agreement with the complete profile obtained by \citet{annette}
from the huge database presented in \citet{mcc}; A. Ferguson, private
communication). The above results may suggest that a weak old-halo component is
indeed present also in M~33, but probably the point could be established only
with an analysis of the stellar kinematics in the considered radial ranges. The
possibility to interpret the observations of the stellar content at large
distance from the center of the galaxy in terms of an unexpectedly metal rich
halo or of a very extended old disc is (again) reminescent of the case of M~31 
\cite[see][for discussions and references]{ann_disc,m31}.

   \begin{figure}
  \centering
   \includegraphics[width=9.3cm]{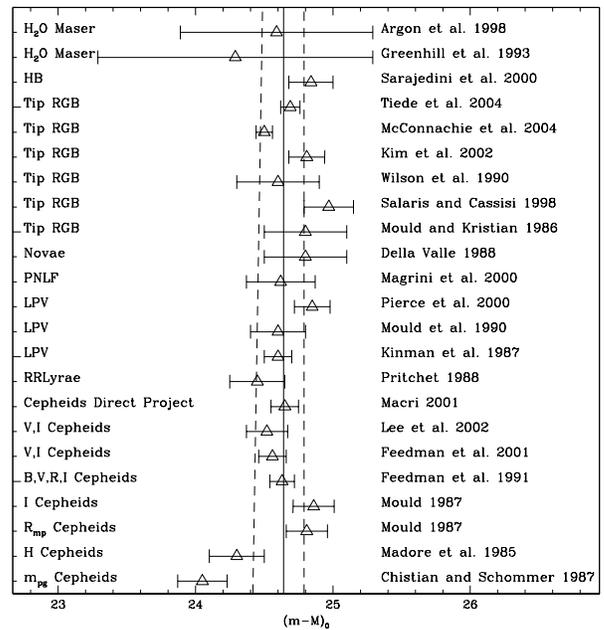}
     \caption{Previous estimates of the distance modulus from the literature
     (open triangles) are compared with the estimate obtained in the present
     study (continuous vertical line). 
     The dashed line encloses the uncertainty
     range of our estimate. The method adopted in the different
     estimate is indicated on the left part of the plot, the sources on the 
     right part. HB = Horizontal
     Branch; PNLF = Planetary nebulae Luminosity Function; LPV = Long Period
     Variables.
               }
          \label{FigDista}
  \end{figure}
%

\begin{acknowledgements}
This research is partially supported by
the Italian {Ministero  dell'Universit\'a e della Ricerca Scientifica}
(MURST) through the COFIN grant p.  2002028935-001, assigned to the project 
{\em
Distance and stellar populations in the galaxies of the Local Group}.
This work  was supported by 
a fellowship (S.G.) from the {\it Consorzio Nazionale Astronomia 
ed Astrofisica--CNAA} and contributions from {\it 
MIUR-COFIN}.
Part of the data analysis has been performed using software developed by P.
Montegriffo at the INAF - Osservatorio Astronomico di Bologna.
This research has made use of NASA's Astrophysics Data System Abstract
Service. The kind assistance of the TNG staff during the observing run is also
acknowledged.

\end{acknowledgements}

\end{document}